# Fast Subsequent Color Iris Matching in large Database

Adnan Alam Khan[1], Safeeullah Soomro[2] and Irfan Hyder[3]

[1] PAF-KIET Department of Telecommunications, Employer of Institute of Business Management
Karachi, Sindh, Pakistan

[2] Institute of Business and Technology, Biztek
Karachi, Sindh, Pakistan

[3] Pakistan Air Force Karachi Institute of Economics and Technology,
Karachi, Sindh, Pakistan

## Abstract

Databases play an important role in cyber world. It provides authenticity across the globe to the legitimate user. Biometrics is another tool which recognizes humans using their physical statistics. Biometrics system requires speedy recognition that provides instant and accurate results. Biometric industry is looking for a new algorithm that interacts with biometric system reduces its recognition time while searching its record in large database. We propose a method which provides an appropriate solution for the aforementioned problem. Iris images database could be smart if iris image histogram ratio is used as its primary key. So, we have developed an algorithm that converts image histogram into eight byte code which will be used as primary key of a large database. Second part of this study explains how color iris image recognition can take place. For this a new and efficient algorithm is developed that segments the iris image and performs recognition in much less time. Our research proposes a fast and efficient algorithm that recognizes color irises from large database. We have already implemented this algorithm in Matlab. It provides real-time, high confidence recognition of a person's identity using mathematical analysis of the random patterns that are visible within the iris of an eye.

## I. INTRODUCTION

Every passing day is bringing innovation in technology and services in today's world. There is a rapid increase in human activities and transaction which eventually requires quick and trusted personal identification. All identification efforts such as, computer login control systems, ATM machines etc., provides reliability, automation and speed. There are a variety of technologies available to effectively identify someone at work, at the gym, at public events, in court houses and even at home. However, producers and buyers of these security tools can not stop thinking about new ways to get the most lucrative contracts from private companies or government. A unique, readily measured and invariant biometric factor for each individual is required for identification purpose usage of biometric indicia. Fingerprints, signatures, photographs, retinal blood vessel patterns and voiceprints have quite a few and significant disadvantages. Though fingerprints and photographs are relatively cheap and offer easy usage but there is no assurance, hence, can easily be manipulated and forged. Similarly, voiceprints, handprints and fingerprints can easily be counterfeited. However, Iris, being an internal organ of human eye, provides automation, speed and reliability to the identification system. It is protected from external environment which makes it a perfect biometric for the identification purpose.

The Purpose of this paper is to focus on the theory behind Iris Recognition System and further experiment and implement this concept. Pattern recognition, optics, statistical inference and computer vision together form Iris recognition technology. Iris, being an internal organ can server to be an automated password which a person does not even have to remember. It offers full confidence recognition for the identity of a person as iris random textures are stable throughout the life time. Frank Bruch, an ophthalmologist, originally provided the concept of using iris patterns for personal identification, in the year 1936.

This system contains no flaws only advantages however but the only disadvantage is the cost of the equipment, smart databases and its backups. Smart database means a database which search iris image from large iris database in fraction of seconds. It means there is requirement to develop an advance database that can provides authentic information as it is required. In this regard we have developed a new algorithm that generates a secret code from iris image using its histogram and near infra-red (NIR) values and assigns a code to that image. This code plays an important role for our recognition system and database indexing; we give it a name which is HNIR. Actually this code is generated by an actual histogram value from an image that's why it got unique value and we use this code as primary key. Second part of our research is to select the best iris image form large database. Image segmentation is another approach that classify image into four segments of left and right irises and compare the result with previously saved image in a database. These iris images are taken in NIR light and contain valuable pattern. This pattern is unique for each individual that's why it is easy to identify a person once it passes through that system. Our contribution is related with database indexing using image HNIR that helps to search nearest pattern from thousands of previously stored images. In other word we can say that it is most secured and accurate biometric system for user authentication. Role of biometric system in our society is quite appreciable. For image analysis and detection iris input plays a key role in this system.

CCD cameras and NIR (Near infrared light) is used for this purpose. NIR is not visible for human eye, and camera can take fine iris images. CCD cameras are composed of multi spectral sensors (e.g. RGB sensors) which offers a lot of advantages. On the other hand, IR camera is that they provide good quality images only in ideal situations which leave a question on its credibility under real conditions [1]. Further, IR cameras are not used commonly and are comparatively more expensive than the colored devices. Color cameras provide richer and multi-dimensional and high performance contents as compared to the IR cameras.

The problem in this case is how to combine the information coming from the different channels of the multi spectral sensors. To deal with this problem, data fusion can represent an efficient solution [2]. Blue channel improves recognition rate in many aspects, if we combine blue channel with any other channel can provide a substantial improvement in performance, and the IR and Red channels performed very well for the brown irises.[3][4][5] Our approach is to take color iris image from specific distance. It detects iris through circle and centralized by sclera corners or edges. Once edges are mapped the iris is locked and zoomed to 235*335 pixels (As shown in Figure2).

Rule of photography is based on exposure value or "ev", i-e the available light or brightness in an image. It will cause related higher brightness in a single pixel or pixel intensity of an image.

In other words we can say that pixel intensity of an iris image is the multiple of MxN and mathematical figure (0.51 to 0.53) as its exposure value.[6][7]

$$\text{Threshold}_{image}\{F(u,v)\} = \frac{1}{M \cdot N_{SIZE}} \sum_{X=0}^{M-1} \sum_{T=0}^{N-1} \text{Intensity} \underset{Pixel}{(f(x,y))}$$

For u=0—>M-1 and v=0—>N-1
$$(0.51 \rightarrow 0.53).$$

Iris localization is another major step it's based on the finding circle in an iris. Locating circle in pupil gave us the center of a circle $(X_0, Y_0)$ in an iris and using its center we find the vertical "$d_v$" and horizontal distance "$d_h$" of the human iris. Product of horizontal and vertical distance gave us the total area "$A_{Pixels}$" for recognition.

Iris is ready to compare with its original image saved in the database. Finding Hamming distance "$H_D$" the basic concepts behind all recognition. Its mathematical expression is as follows.

$$\text{Hamming Distance } H_D = \frac{(A \oplus B) \otimes (\text{Actual Mask})}{(\text{No of Bits})_{TEMPLATE} - (\text{No of Bits})_{ACTUAL \; MASK}}$$

## 2. Methodology

Iris recognition is the one of an efficient approach in biometric identification. Color Iris Recognition System CIRS is based on histogram generated codes and pattern matching. In this regards we use two types of data sample standard database UBIRIS.v1 and manual iris samples. CIRS works on control conditions it means it requires near infra light cameras, proper lightening, and standard computer which runs this algorithm. This algorithm has two separate mode manual recognition and auto detection. In manual recognition user must select filter values, color signal mathematical operations at its own. Where else auto system choose the best at its own. Mathematical relationship for are as follows.

$$F_R(x', y') = \int Cp(x', y', \lambda_1) \, V_R(\lambda_1) \, d\lambda_{1_\ast}$$

$$F_G(x', y') = \int Cp(x', y', \lambda_2) \, V_G(\lambda_2) \, d\lambda_2$$

$$F_B(x', y') = \int Cp(x', y', \lambda_3) \, V_B(\lambda_3) \, d\lambda_{3_\ast}$$

Aforementioned equations depict the mathematical equations of different wavelengths mainly Red, Green and Blue. These wavelengths generally denoted by $\lambda$ and nanometer is the unit of these wavelengths. [3]

The size of a color image matrix is three times the size of monochrome image matrix. This smart database saves iris images; histogram based unique code and authorized user data. It contains a lot of user information and sample iris images.

Here we choose near infra-red light images which provides clear iris pattern with less reflections which is suitable for iris recognition system. This is the first part of research. Second part of this project is key generation using image histogram which will saves as primary key of image database further its used as recognition key . Third part of this research is database connectivity, we have developed this project in Matlab, it has another unique feature which is it connectivity with other databases. Fourth part of this project is storage; here data is stored in the form of images not text or special characters. Fifth database security, Matlab provides security to its own built in database so data cannot be vulnerable. Sixth part is the efficiency; our algorithm compares the small portion of an iris that's why it takes minimum time for recognition.

Color Iris Recognition System CIRS algorithm based on the following steps.

| | | | |
|---|---|---|---|
| STEP 0 | START | STEP 26 | SET |
| STEP 1 | LOOP COMMENTS image tagging | STEP 27 | PRIMARY KEY-->CODE Image RGB |
| STEP 2 | ADD IRIS image RGB LEFT/RIGHT | STEP 28 | INDEXED DATABASE |
| STEP 3 | LOAD image RGB | STEP 29 | CLOSE DATABASE |
| STEP 4 | READ image RGB | STEP 30 | END LOOP |
| STEP 5 | CLASSIFY image RGB | STEP 31 | END |
| STEP 6 | image R; image G; image B | STEP 32 | START |
| STEP 7 | ADJUST image R | STEP 33 | LOOP COMMENTS iris image recognition |
| STEP 8 | CALCULATE RED/image R | STEP 34 | LOAD X-image RGB |
| STEP 9 | SAVE R | STEP 35 | LOAD DATABASE |
| STEP 10 | ADJUST image G | STEP 36 | GENERATE X-image RGB code |
| STEP 11 | CALCULATE GREEN/image G | STEP 37 | CALCULATE nearest image RGB |
| STEP 12 | SAVE G | STEP 38 | SAVE nearest image RGB |
| STEP 13 | ADJUST image B | STEP 39 | RUN image RGB RECOGNITION |
| STEP 14 | CALCULATE BLUE/image B | STEP 40 | LOAD LEFT/RIGHT image RGB |
| STEP 15 | SAVE B | STEP 41 | CLASSIFY image RGB |
| STEP 16 | READ image visible threshold V | STEP 42 | ONE/TWO/THREE/FOUR |
| STEP 17 | READ image Near infrared threshold NIR | STEP 43 | CHOOSE any number |
| STEP 18 | GRAPH V,NIR | STEP 44 | UNION/INTERSECTION |
| STEP 19 | CALCULATE SUBTRACT (NIR,V)/ADD(NIR,V) | STEP 45 | CHOOSE U/I |
| STEP 20 | SAVE ratio | STEP 46 | GENERATE CODE alpha |
| STEP 21 | LOOP ratio>x | STEP 47 | COMPARE CODE (aplha,DATABASE) |
| STEP 22 | CALCULATE Area(ratio) | STEP 48 | CALCULATE nearest image RGB |
| STEP 23 | END LOOP | STEP 49 | SHOW RESULTS |
| STEP 24 | CONCAT (R,G,B,ratio)-->CODE Image RGB | STEP 50 | END LOOP |
| STEP 25 | OPEN DATABASE | STEP 51 | END |

Figure1: The Algorithm

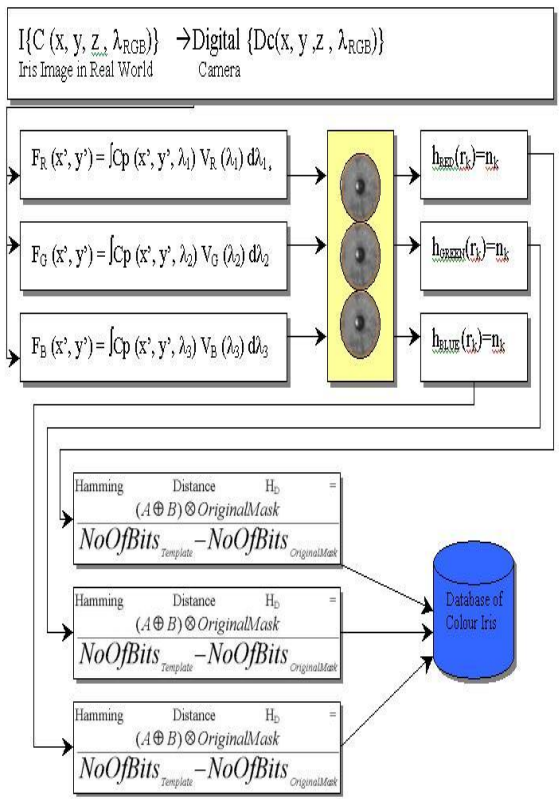

Figure2: Related Mathematics

This study explains that color iris recognition is possible in large databases in few conditions. First assigning histogram Red/Green/Blue and NIR code with that image as an iris image ID. Secondly pattern matching algorithm can identify the exact iris image. In other words study proves that HNIR as a primary key can recognize an image due to its histogram characteristics and latter it confirms it using pattern matching algorithms. Interior ministry of any country can use this smart algorithm to identify its citizens or culprits easily. This algorithm is developed for large collection of records and its main use to identify iris to iris directly rather name or NIC. Our proposed Matlab code converts color image into separate red, green, blue signals and calculates its histogram. This generated histogram is combined with another histogram of Near Infrared light and provides a unique code. This code is used as primary key of color iris image database. This database is separately designed for these types of searches. Fields of this database contains name of the authorized person, his national ID, address, family records. The most important part of this project is the six separate unique codes matching; it means a color iris image can produce six separate unique codes after histogram matching. Actually these codes are calculated by the system and matched it with newly input image. It means system sends codes to match with newly input image. Here two simultaneously processes run one with the database and with the input image for recognition. Let say if the histogram result is matched but green

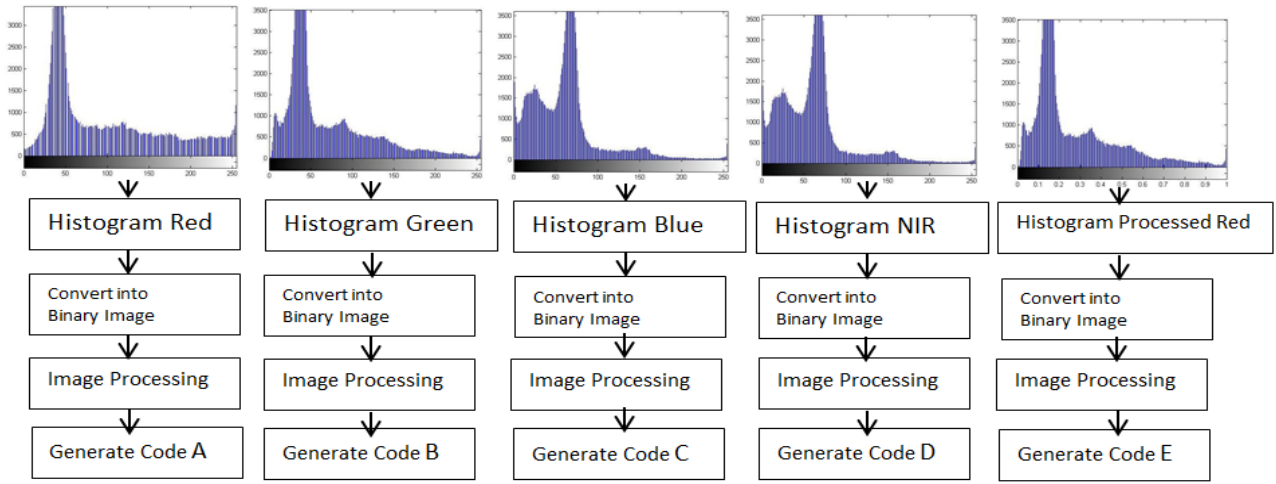

Primary Key (PK) =Iris image { CodeA||CodeB||CodeC||CodeD}
IRIS DATABASE=Indexed {Primary Key (PK)}
Search (IRIS DATABASE) Results= Many [Results {Iris image}]
Unique IRIS Result= CIRS (Many [Results {Iris image}])

Figure3: Code Generation

pattern matching is not matched so user can change the signal from green or blue to red. Its results are more clear and visible.

Iris image enhancement is one of the most important phenomena for iris recognition. Iris image filtering is one of the most important methods that can improve the recognition rate because it clears the iris image pattern. Here we use different image enhancement filters mainly imtophat filter etc.

## 3. Results

For this study UBIRIS.v1 is used and we picked of fifty different samples to test our proposed algorithm. It results are as follows.

| Recognition Color | 235*335 Focused | 235*335 Not Focused |
|---|---|---|
| Red | 91% | 10% |
| Blue | 53% | 5% |
| Green | 76% | 7% |

Histograms selects the nearest class images and shows the results that these images are near to this sample than we run the second part of the algorithm which choose the best among best. Our study proves that our proposed method takes less computational time than the latest color iris recognition system. Image segmentation reduces ¼ average recognition rates. Its accuracy is discernible because it matches biometric code with self-developed code using different mathematical operations.

## 4. Conclusion

Iris recognition is a method of biometric authentication that uses pattern recognition technique based on high-resolution color iris images of an individual's eyes. The aim of this project is to design a smart database that matches image with image using histogram and shows nearest iris classes or images. Further it runs pattern matching algorithm which selects the one quadrant out of four and generate code using RGB signals. Proposed color iris recognition methodology for complex system in controlled environment is efficient than other systems. Mathematical working and its explanations are focused in this paper. This Iris recognition system takes 75% less time from other color iris recognition system.

## 5. References

[1] Caitang Sun, Farid Melgani, Chunguang Zhou,De Natale Francesco, Libiao Zhang, Xiangdong Liu, "Semi-Supervised Learning Based Color Iris Recognition", *Proc. of the Fourth International Conference on Natural Computation*,2008.

[2] Caitang Sun, Farid Melgani, Chunguang Zhou,De Natale Francesco, Libiao Zhang, Xiangdong Liu, "Incremental Learning based Color Iris Recognition", *Proc. of 2008 Seventh Mexican International Conference on Artificial Intelligence*,2008.

[3] Onur G. Guleryuz, "Image Formation" *Department of Electrical and Computer Engineering, Polytechnic University, Brooklyn, NY 1*,2003.

[4] J. Daugman, R. Wildes, W. Boles, "Statistical decision theory of Iris", *IEEE*, 2004.

[5] Emine, Sonia, Bernadette ,"Iris Identification Using Wavelet Packets", *Proc .of 17th International Conference on Pattern Recognition (ICPR'04)1051-4651/04*, IEEE, 2004.

[6] Hanho Sung, Jaekyung Lim, Ji-hyun Park, Yillbyung Lee, "Iris Recognition Using Collarette Boundary Localization",*17th International Conference on Pattern Reg (ICPR'04),1051-4651/04,* IEEE,2004.

[7] Li Ma, Yunhong Wang, Tieniu Tan, "Iris Recognition Using Circular Symmetric Filters" *proc .of 16th International Conference on Pattern Recognition (ICPR'02) 1051-4651/02* IEEE, 2002.